\documentclass{llncs}

\usepackage{amsmath}

\usepackage{compozition}

\setdefaultarrow{stealth}

\begin{document}

\title{On The Theory of Composition in Physics}
\titlerunning{Theory of Composition}
\author{Lucien Hardy}
\institute{Perimeter Institute, 31 Caroline Street North, \\ Waterloo, Ontario N2L 2Y5, Canada}
\date{}

\maketitle

\begin{abstract}
We develop a theory for describing composite objects in physics. These can be static objects, such as tables, or things that happen in spacetime (such as a region of spacetime with fields on it regarded as being composed of smaller such regions joined together).  We propose certain fundamental axioms which, it seems, should be satisfied in any theory of composition.  A key axiom is the order independence axiom which says we can describe the composition of a composite object in any order.   Then we provide a notation for describing composite objects that naturally leads to these axioms being satisfied.  In any given physical context we are interested in the value of certain properties for the objects (such as whether the object is possible, what probability it has, how wide it is, and so on).  We associate a \emph{generalized state} with an object.   This can be used to calculate the value of those properties we are interested in for for this object.   We then propose a certain principle, \emph{the composition principle}, which says that we can determine the generalized state of a composite object from the generalized states for the components by means of a calculation having the same structure as the description of the generalized state.  The composition principle provides a link between description and prediction.
\end{abstract}


\section{Introduction}

It is a great pleasure to contribute to this festschrift for Samson Abramsky.   Over the last twenty years the worlds of quantum theory and computer science have collided giving rise to quantum information and quantum computing.   What has singled out Samson's approach has been the emphasis on fundamental structure.  Physicists tend to rely on partial differential equations and simple circuit models.  Computer science, on the other hand, investigates a much richer and deeper set of paradigms for dynamics.  Samson has been the leading force in bringing these deeper structural insights to bear on quantum theory through his work on the use of category theory in quantum information \cite{abramsky2004categorical,abramsky2007physics}, his work on investigating the sheaf-theoretic structure of various no-go theorems in quantum foundations \cite{abramsky2011sheaf,abramsky2012logical}, and much more (for example \cite{abramsky2012relational,abramsky2012operational,abramsky2010relational,abramsky2011big,abramsky2007petri,abramsky2007temperley}).  The present contribution is in the same spirit, and Samson's influence will be clear to all readers of this volume.

Science proceeds by analyzing big things in terms of smaller things while keeping note of how these smaller things are joined together. In other words, it concerns composite objects.  The composite objects we wish to analyze may be static objects to be specified at a given time or they may be something that happens in spacetime.  Thus, a table is made out of pieces of wood joined together appropriately.  A region of spacetime with fields on it can be thought of as being composed of many smaller regions of spacetime joined at their boundaries.  A quantum optical experiment can be thought of as consisting of many apparatus uses (or operations) joined together so that apertures on these apparatuses are aligned. And so on. In many different branches of physics (and indeed science generally) we see ideas of composition.  Indeed, it is difficult to imagine scientific explanation that does not correspond to some kind of compositional analysis.  The usual approach is to reinvent basic ideas about how these smaller parts are composed to make bigger things every time we set up a new physical theory.  It makes sense to think about composition of this sort in the abstract.  Having obtained general notions about the theory of composition, we can see how these work for particular applications.  In this way we may gain useful insight about how science works in general.  This is the subject of this paper.

In the example of a table we can describe how different parts are actually joined (by somebody) to make the table.  However, we do not need to imagine that somebody is actually putting the composite object together for a theory of composition to be useful.  Rather, we may merely be analyzing an object in terms of its parts.  This is the case in the example of a region of spacetime with fields on it.  Given some larger region we can divide it up into smaller regions (in many different ways).  Nobody is actually building the larger region out of smaller regions.  There may, indeed, be many ways of regarding a bigger object as being made out of smaller parts.

Initially we will be interested in the \emph{description} of composite objects irrespective of whether some particular laws of physics actually allow any given object to exist.  Indeed, an interesting example of a composite object that we can describe, yet is impossible, is the Penrose (or impossible) triangle.
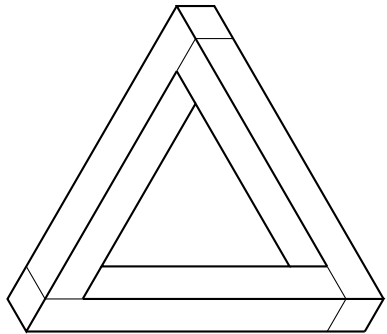
\[
\begin{Compose}{0}{0}
\draw[thick] (0,0) coordinate(A) -- (10,0) coordinate (B) -- (5, 8.66) coordinate (C) -- cycle;
\draw[thick] (A) -- ++ (-1, -1.732) coordinate (AA); \draw[thick] (B) -- ++ (2, 0) coordinate (BB); \draw[thick] (C) -- ++(-1, 1.732) coordinate (CC);
\draw[thick] (AA) -- ++(16, 0) coordinate (BBB) -- ++ (-120:2) coordinate (BBBB);
\draw[thick] (BB) -- ++ (120: 16) coordinate (CCC) -- ++(2,0) coordinate (CCCC) -- (BBB) ;
\draw[thick] (CC) -- ++(-120: 16) coordinate (AAA)-- ++(120:2) coordinate(AAAA) -- (CCC);
\draw[thick] (BBBB) -- (AAA);
\draw[thin] (BB) -- (intersection cs: first line={(CCC)--(BB)}, second line={(AA)--(BBB)} ) coordinate (B5) -- ++($(BBBB) - (BBB)$);
\draw[thin] (AA) -- (intersection cs: first line={(BBB)--(AA)}, second line={(CC)--(AAA)} ) coordinate (A5) -- ++($(AAAA) - (AAA)$);
\draw[thin] (CC) -- (intersection cs: first line={(AAA)--(CC)}, second line={(BB)--(CCC)} ) coordinate (C5) -- ++($(CCCC) - (CCC)$);
\end{Compose}
\]
This well known example consists of three straight beams each having square cross section and each legally joined to the other two at the ends at right angles (as shown).  The resulting object, however, is clearly impossible.  In our theory of composition we would be able to describe this object simply by specifying the three objects and the way they are joined.

Having discussed the description of composite objects, we will then go on to discuss what \emph{predictions} different physical theories might make about such objects.  The sort of statements a physical theory will make depends on what type of physical theory we have.  Deterministic theories can give us a yes or no answer to whether a particular object can exist.  Probabilistic theories might give us a probability.    We will propose a principle that may be of some use in setting up new physical theories.  This is the  \emph{composition principle}.   This principle says that we can establish the generalized state for a composite object by means of a calculation having the same structure as the description of the composition of that object.

In this paper we will be concerned with some preliminary ideas concerning composition.  We will set up some further notation and introduce some axioms for composition which may, or may not, be satisfied in certain situations. Our original motivation for these considerations comes from quantum theory.  However, the claim is a stronger one. Namely that all physical theories can be understood in these terms and that the composition principle can always be made to hold.  The hope is that these ideas can play a role in the construction of a theory of quantum gravity.  With regard to quantum theory it is possible that, by thinking about composition in an abstract and structural way, we may be able to provide deeper motivation for the kind of axioms that have recently been used in operational reconstructions of quantum theory \cite{hardy2001quantum,dakic2009quantum,masanes2010derivation,chiribella2010informational,zaopo2012information,hardy2011reformulating}.  In particular, we will see that there is a link between the axiom of tomographic locality and the composition principle proposed here.

\section{Examples and The Tensorial Notation}\label{SecTables}

\subsection{Tables}

To get us warmed up let us consider an example.  A piece of furniture is composed out of many separate parts joined together in different ways.  For example, the following (two-dimensional) table
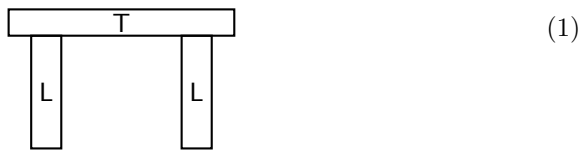
\begin{equation}
\begin{Compose}{0}{0}
\Crectangle{T}{6}{0.7}{0,0} \Crectangle{L}{0.8}{3}{-4,-3.7} \crectangle{L2}{0.8}{3}{4,-3.7} \csymbol{L}
\end{Compose}
\end{equation}
is made by joining legs, $\mathsf L$, to a top $\mathsf T$.  We can illustrate how the table is put together by showing the joins as follows
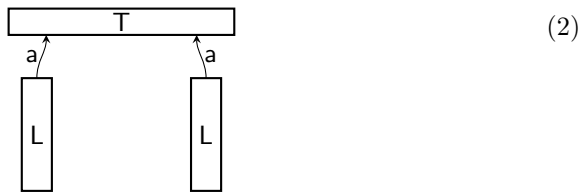
\begin{equation}\label{tabelillustration}
\begin{Compose}{0}{0}
\Crectangle{T}{6}{0.7}{0,0} \Crectangle{L}{0.8}{3}{-4.5,-6} \crectangle{L2}{0.8}{3}{4.5,-6} \csymbol{L}
\jointb[left]{L}{0}{T}{-4} \csymbol{a}  \jointb[right]{L2}{0}{T}{4} \csymbol{a}
\end{Compose}
\end{equation}
Here we join the leg to the table by a join of a type we describe as $\mathsf a$.  We have an arrow because joins are typically asymmetric. For example, we may join the leg to the table by putting glue on the leg then fixing it in place on the table (a stronger join may involve screws as well).   In this illustration we retain the shapes of the objects.  We can use more abstract notation
\begin{equation}
\begin{Compose}{0}{0.5}
\Ucircle{T}{0,0} \Ucircle{L}{-3,-4}  \ucircle{L2}{3,-4} \csymbol{L}
\joincc[above left]{L}{90}{T}{-110} \csymbol{a}
\joincc[above right]{L2}{90}{T}{-70} \csymbol{a}
\end{Compose}
~~~~ \Leftrightarrow ~~~~ \mathsf{L^{a_1}L^{a_2} T_{a_1a_2} }
\end{equation}
We will call this abstract notation \emph{the tensorial notation} because of its similarity with tensor notation (though, of course, here it is being used to describe composite objects).
The advantage of more abstract notation is that we may use the same notation across different branches of physics (not just furniture).  In the diagrammatic notation (on the left) we represent each object by a circle.  The table has two different positions one can join legs and these are represented by the arrow entering the circle in different places.  In the symbolic notation the two different positions are represented by whether we consider the first or second subscript respectively on $\mathsf{T_{a_1a_2}}$.  In the symbolic notation we need to introduce integers, $1$ and $2$, to label the different joins.  These integers have no significance beyond that they label the joins.  We can relabel them with different integers (or permute the integers used) and have the same object.

The tensorial notation carries no particular order on the objects.  For example, we can write
\begin{equation}
\mathsf{L^{a_1}L^{a_2} T_{a_1a_2} = L^{a_2} T_{a_1a_2}L^{a_1}= T_{a_1a_2}L^{a_1}L^{a_2}=L^{a_2} L^{a_1}T_{a_1a_2} }
\end{equation}
This lack of order is even clearer in the diagrammatic notation.  There we do not care where the circles are placed on the page so long as we preserve the exit and entry positions on the circles for the joins as well as preserving the graphical information.  For example
\begin{equation}
\begin{Compose}{0}{0}
\Ucircle{T}{0,0} \Ucircle{L}{-3,-4}  \ucircle{L2}{3,-4} \csymbol{L}
\joincc[above left]{L}{90}{T}{-110} \csymbol{a}
\joincc[above right]{L2}{90}{T}{-70} \csymbol{a}
\end{Compose}
=
\begin{Compose}{0}{0}
\Ucircle{T}{0,0} \Ucircle{L}{-6,2}  \ucircle{L2}{-2,-4} \csymbol{L}
\joincc[below left]{L}{90}{T}{-110} \csymbol{a}
\joincc[below]{L2}{90}{T}{-70} \csymbol{a}
\end{Compose}
\end{equation}
This lack of order is important.  Often we are tempted to think in terms of there being a particular order in particular physical situations.  For example, when we buy a piece of self-assembly furniture it usually comes with detailed instructions for what order one is advised to put the pieces together in.  This is useful advice to have if one is trying to build a piece of furniture.  However, here we are interested simply in describing a composite object, not in accounting for how it might have been put together.  Indeed, there may be many alternative orders in which to assemble a given piece of furniture.   This idea that one should not think of the objects in some particular order will play an important role in what follows.

\subsection{Circuits}\label{SecCircuits}

A common type of situation in physics is a circuit (such as in quantum theory).  A circuit consists of operations wired together. Here is an illustration of a circuit:
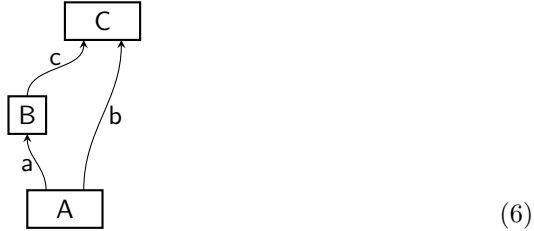
\begin{equation}
\begin{Compose}{0}{0}
\Crectangle{A}{2}{1}{0,0} \Crectangle{B}{1}{1}{-2,5} \Crectangle{C}{2}{1}{2,10}
\jointb[left]{A}{-1}{B}{0} \csymbol[0,-4]{a} \jointb{B}{0}{C}{-1} \csymbol{c} \jointb[right]{A}{1}{C}{1} \csymbol[5,0]{b}
\end{Compose}
\end{equation}
which can be described by the notation
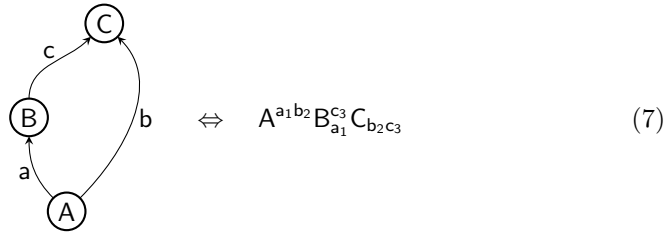
\begin{equation}
\begin{Compose}{0}{-1.3}
\Ucircle{A}{0,0} \Ucircle{B}{-2,5} \Ucircle{C}{2,10}
\joincc[left]{A}{135}{B}{-90} \csymbol[0,-4]{a} \joincc{B}{90}{C}{-135} \csymbol[-3,3]{c} \joincc[right]{A}{45}{C}{-45} \csymbol[5,0]{b}
\end{Compose}
~~~ \Leftrightarrow~~~ \mathsf{ A^{a_1b_2} B_{a_1}^{c_3} C_{b_2c_3}}
\end{equation}
An operation corresponds to one use of an apparatus.  The apparatus has apertures that are aligned with each other so that systems of various types, denoted by $\mathsf a$, $\mathsf b$, etc can pass between the operations.  A system of type $\mathsf a$ passing from one operation to another corresponds to a join which we will say is of type $\mathsf a$.   The apparatus has on it outcomes (such as lights that flash, a meter whose needle points to a particular number, a LED display indicating an outcome, etc).  The operation has a subset of all the possible outcomes associated with it (this is implicit in the symbol $\mathsf A$).  We will say that the operation \lq\lq happens" if the outcome recorded is in the set of outcomes associated with the given operation.  In a deterministic theory a given circuit is either possible or impossible.  In a probabilistic theory a given circuit will have a probability for happening (the probability that the outcome at each circuit is in the set of outcomes for that operation).  A general operational theory for circuits (with particular application to quantum theory) using the notation in this paper is given in \cite{hardy2010formalism,hardy2011reformulating,hardy2012operator}.

\subsection{Other Examples}

We can use the same notation to describe many different kinds of situations. Here are some examples:
\begin{description}
\item[Penrose objects.]  This is the class of objects described in the following way. We take a number of square cross-section beams of a certain length.  Then each beam can be joined to others at right angles at its ends.  Most objects so formed will be impossible.  In particular, we can describe Penrose triangles and generalizations thereof (such as Penrose squares, Penrose pentagons, \dots) in this notation. Some objects described in this way will be possible.
\item [Spacetime with fields.] We can describe a region of spacetime with fields defined on it that is being regarded as composed of many joined smaller regions.
\item[Statics problems.] We can describe statics problems (such as a ladder leaning against a wall).
\item[Distances.] We can describe as composite a region of spacetime in which every path has a distance.
\item[Minimum distances.] We can describe as composite a region of spacetime in which every pair of points has a minimum distance measured between them along a path in the region
\end{description}
This framework is for the \emph{description} of composite objects without saying anything about whether these objects are possible. In physics we are interested in making predictions.  We may wish to predict whether some particular object is possible, what the probability is for it, or what the value of some property (such as the width) is.  For this purpose we associate a generalized state with an object.  The generalized state can be used to calculate the value of those properties we are interested in for the given object.   A natural question is how do we obtain the generalized state for a composite object?  The \emph{composition principle} discussed in Sec.\ \ref{SecCompositionprincple} is particularly useful in this respect.  It states that we can determine the generalized state for a composite object from the generalized states for the components by means of a calculation that has the same form as the description of the given composite object.  For example,
\begin{equation}
\mathsf{ A^{a_1b_2} B_{a_1} C_{b_2} } ~~\text{has generalized state}~~ A^{a_1b_2} B_{a_1} C_{b_2}
\end{equation}
where $A^{a_1b_2}$, $ B_{a_1}$, and  $C_{b_2}$ are the generalized states associated with the components.  In the case of probabilistic circuits, for example, these mathematical objects are tensors. The placement of subscripts and superscripts indicates some particular mathematical procedure for completing the calculation (it need not be a tensor calculation).  If we can always write down our calculations in this way then the composition principle may turn out to be a very powerful way of obtaining new physical theories.

\section{Background}

A decade or so ago, Abramsky and Coecke initiated a hugely influential approach to quantum theory based on category theory \cite{abramsky2004categorical}.  This emphasizes the compositional structure of quantum circuits and the connection between this compositional structure and the structure of quantum calculations.  One aspect of this work is the use of diagrammatic notation.  This is taken to heart here.  Another tradition in quantum foundations has been the convex probabilities approach (in which states are represented by lists of probabilities).  This approach goes back to Mackey \cite{mackey1963mathematical} and has been worked on by many people over the years including Ludwig \cite{ludwig1985axiomatic}, Davies and Lewis \cite{davies1970operational}, Gunson \cite{gunson1967algebraic}, Mielnik \cite{mielnik1969theory}, Araki \cite{araki1980characterization}, Gudder {\it et al.\ } \cite{gudder1999convex}, Foulis and Randall \cite{foulis1979empirical}, Fivel \cite{fivel1994interference} as well as more recent incarnations \cite{hardy2001quantum,barrett2007information}.  More recently there has been some work combining the diagrammatic approach of Abramsky and Coecke with the the convex probabilities approach (for example, see the papers of Chiribella, D'Ariano, and Perinotti \cite{chiribella2010probabilistic,chiribella2010informational} and the duotensor formulation due to the present author \cite{hardy2010formalism,hardy2011reformulating,hardy2012operator}).  In this paper, the approach developed in \cite{hardy2010formalism,hardy2011reformulating,hardy2012operator} which concerns probabilistic circuits is taken as a springboard to a more general theory of composition.  The notation used there (though in a more restricted context) is the same as the tensorial notation for describing composite object used here.   Thus, rather than writing $\mathsf{(A\otimes B)\circ (C\otimes D)}$ or $\mathsf{(A\circ C)\otimes (B\circ D)}$ for 
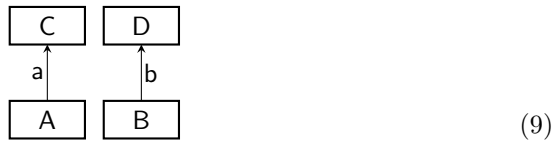
\begin{equation}
\begin{Compose}{0}{0}
\Crectangle{C}{2}{1}{0,5} \Crectangle{D}{2}{1}{5,5}
\Crectangle{A}{2}{1}{0,0} \Crectangle{B}{2}{1}{5,0}
\jointb[left]{A}{0}{C}{0} \csymbol{a} \jointb[right]{B}{0}{D}{0} \csymbol{b}
\end{Compose}
\end{equation}
we write
\begin{equation}
\mathsf{A^{a_1} B_{a_1} C^{b_2} D_{b_2} }
\end{equation}
This change of notation signals a change of attitude.  The tensor product $\otimes$ is replaced by the notion of a \emph{null join} and is treated as yet another type of join, albeit with special properties (as will be discussed in Sec.\ \ref{SecNulljoins} and Sec.\ \ref{SecMoreonNulljoins}). The superscript/subscript structure now tells us where the components are joined. This is a more versatile notation than the $\circ$ symbol which plays the same role in the example given.  This notation is such that we can mix up the order and still denote the same circuit.  For example, $\mathsf{ C^{b_2}B_{a_1} D_{b_2}A^{a_1} }$ corresponds to the same circuit.  This notation is also good for describing other kinds of composite object (such as tables) where we are not concerned with process or the passage of systems between boxes.

Although the ideas in this paper are categorical in spirit, no attempt will be made to bring the formal apparatus of category theory to bear on theory of composition proposed here. However, given unifying power of categorical approaches, it is to be expected that much good would come of such an endeavor.

\section{The Description of Composite Objects}

In Sec.\ \ref{SecTables} we provided a particular notation (which we dubbed the tensorial notation) for describing composite objects.  This notation has, built into it, various assumptions about the nature of composition.  In this Sec.\ \ref{SecBipartitenotation} we wish to take a step back and analyze some of these basic assumptions.  We will do this by suggesting some fundamental axioms. We do not claim that these constitute a complete set of axioms.  They are simply proposed to help us to gain some insight into the notion of composition. The first, and most fundamental, of these is the \emph{composition axiom}. This will allow us to use a more primitive form of notation (than the tensorial notation) which we will call \emph{the bipartite notation}.  Once we have the bipartite notation we propose two more fundamental axioms. These are the \emph{order independence axiom} and the \emph{null joins axiom} (actually a set of axioms concerning null joins).  These axioms take us much, but not all of the way to being able to use the tensorial notation. In Sec.\ \ref{SecTensorialnotation} we will state the composition locality axiom which, basically, is an axiom saying that we can use the tensorial notation for describing composite objects. We will also state the \emph{$R$-enablement axiom} which enables us to regard joins in either direction.  The remainder of this section will deal with joins. We will discuss sufficient sets of joins, and a certain \emph{boundary axiom}.

All these axioms are motivated by thinking about composite objects such as a piece of furniture or a region of spacetime broken up into smaller regions. None of these axioms need be true and some are more basic than others. In this section we are interested in the description of composite objects irrespective of whether the laws of physics say the given object is possible.  We will address what the laws of physics say is possible in latter sections.

\subsection{The Bipartite Notation and Fundamental Axioms}\label{SecBipartitenotation}\label{SecNulljoins}

We assume we have various types of object,  $\mathsf A$, $\mathsf B$, $\mathsf C$, \dots.  We could have more than one object of a given type. An object is fully specified if we provide a specification such that no other object has the same specification.  Objects whose full specifications are the same are of the same type and should be represented by the same letter.   Objects whose full specifications are different are different object types and should be represented by a different letter.  We now state the most basic of axioms.
\begin{description}
\item[Composition axiom:]  Any object composed of two objects is fully specified if one is provided with a full specification of the two component objects and a description of the way in which they are joined.
\end{description}
This axiom corresponds to a kind of reductionism.  Were it not true it would be difficult to develop a theory of composition.  We will denote the ways in which one object can be joined to another by $\alpha$, $\beta$, $\gamma$, \dots.  If we join $\mathsf B$ to $\mathsf A$  by means of join $\alpha$, according to the composition axiom, we fully specify the resulting object by
\begin{equation}
\mathsf{(A,B)_{\alpha}}.
\end{equation}
We call this \emph{the bipartite notation}. Note that the $\alpha$ indicates not only the method of joining but also the join positions on the two objects.  For example, $\mathsf A$ may represent a wooden die, $\mathsf B$ a metal die.  Then $\alpha$ may represent the join in which we attach the six side of a wood die to the five side of the metal die by putting glue on the six of the wooden die and pressing it into place on the five side of the metal die.  Note that there are two aspects of the join.  First, there is the type of join (involving a certain area and glue) and then there is the position of the join on the two objects.  We will denote the type of join by $\mathsf a$, $\mathsf b$, \dots and the position of the join by $x$, $y$, \dots. Thus, we have $\alpha=(\mathsf{a}, x, y)$ where $\mathsf a$ is the type of join (applying glue over a certain area) and $x$ is the position of the join on the first object (the six side of the wooden die) and $y$ is the position of the join on the second object (the five side of the medal die).  Another example is in a circuit.  Then a join $\alpha=(\mathsf a, x, y)$ would correspond to aligning the aperture at position $x$ on the first operation with the aperture at position $y$ on the second operation so that a system of type $\mathsf a$ can pass between the two operations.

Joins are, in general, asymmetric. For example, we put glue on the first object then press it into place on the second object.  We define $\alpha^{\mathnormal R}$ through
\begin{equation}
\mathsf{(A,B)_{\alpha} = (B, A)_{\alpha^{\mathnormal R}}}.
\end{equation}
Hence, if $\alpha=(\mathsf a, x, y)$ then $\alpha^{\mathnormal R}=(\mathsf a^{\mathnormal R}, y, x)$.  Here $\mathsf a^{\mathnormal R}$ is the same join type as $\mathsf a$ but described in the reverse direction (for example, we put glue on the second object then press it into place on the first object).

Here are some examples of composite objects.
\begin{equation}
\mathsf{ ((A, B)_{\alpha}, C)_{\beta}  } ~~~ \mathsf{ ((A, B)_{\alpha}, C)_{\beta}, ~~~ ((A,B)_{\alpha}, ((C,A)_{\beta}, A)_{\gamma})_{\delta}, ~~~
(((A, B)_{\alpha}, C)_{\beta}, D)_{\delta}  }
\end{equation}
We get quite complicated bracketting structure here indicating a particular order of composition.


 The second most fundamental axiom we propose is the following
\begin{description}
\item[{\bf Order independence axiom:}] A given composite object with three components can be regarded as being composed in any order.  Thus, if
\[\mathsf{ D=(A, (B, C)_\beta)_\alpha}\]
then there exists some $\gamma$, $\delta$, $\mu$, $\nu$ such that
\[\mathsf{ D=((A, B )_\gamma, C)_\delta}  ~~~ \text{and} ~~~ \mathsf{ D=((A, C)_\mu, B)_\nu} \]
\end{description}
This is a very basic axiom since, were it not true, it would matter what order we chose to describe the way in which the composite object is built up from its components.  However, if we take this axiom to heart it has potentially far reaching consequences since, in fact, we do typically choose a special order when analyzing composite objects for some physical situations.  The most pertinent example is objects composed in time.  For example, we may evolve a state through a number of discrete time steps.  Typically we consider the corresponding patches of space time in sequence.  The point of the order independence axiom in this case is that one can consider the components of such physical objects in any order, not just the order suggested by the sequence in time.

Whichever way we choose to divide a composite object into two parts, these two parts are, by definition, joined.  However, one way of joining is where we do not directly join two things but merely consider them both to be \lq\lq part of the picture" (and specify no further relationship between them).  We represent this by $\mathsf (A, B)_0$ and call $0$ the null join. We will say that two objects joined by the null joint are \emph{disjoint}. It is reasonable to assume this particular way of joining objects will have special properties since otherwise it would not be distinguished from other types of join.  We assume, then, that
\begin{description}
\item[{\bf Null joins axioms}] Null joins have the following properties:
\begin{description}
\item[Universality:] any pair of objects, $\mathsf A$ and $\mathsf B$, can be joined by a null join.
\item[Uniqueness:]   If $\mathsf{(A,B)_\alpha = (A,B)}_0$ then $\alpha=0$.  
\item[Symmetry:]  We have $\mathsf{(A,B)}_0 = \mathsf{(B,A)}_0$ for any pair of objects, $\mathsf A$ and $\mathsf B$.
\item[Refinement:] if an object is joined to a composite object by the null join then it is appended to each component of the composite object by the null join.
    This means that if we have an object $\mathsf{ ((A, B)_\alpha, C)}_0$ and we write $\mathsf{ ((A, B)_\alpha, C)}_0 = \mathsf{((A,C)_\beta, B)}_\gamma$ (by the order independence axiom) then we must have $\beta=0$.
\end{description}
\end{description}
Although we give the null join a special name and propose a certain axioms for it, we do not otherwise treat it on a different footing to other joins.  It is just another type of join.  If we are going to compose the space time regions associated with three sequential time steps in an order that is not that of the sequence then we might use the null join.  Thus, if these three time steps are $\mathsf A$, $\mathsf B$, and $\mathsf C$ to be taken in that sequence then we could write $\mathsf{ ((A, B)_\alpha, C)_\gamma}$ where the join $\alpha$ is that between the first and second time step and the join $\gamma$ is that of attaching the third time step.  But we might write $\mathsf{ ((A, C)_0, B)_\delta}$ where $\delta$ is the join of inserting the time step $\mathsf B$ in between $\mathsf A$ and $\mathsf C$.  It is reasonable to expect that physical properties for objects connected by the null join will behave in certain ways. For example, in the duotensor framework \cite{hardy2010formalism,hardy2011reformulating,hardy2012operator}, the probability for a circuit composed of two disjoint parts factorizes.

\subsection{Tensorial Notation and Composition Locality}\label{SecTensorialnotation}

We will now take steps taking us from the bipartite notation, $\mathsf{ (A, B)_\alpha}$, to the tensorial notation in the light of the two fundamental axioms just introduced.  The bipartite notation is good for describing a bipartite composite object. However, when we consider an object consisting of more than two parts it becomes cumbersome. A tripartite object is represented by $\mathsf{ ((A,B)_\alpha, C)_\gamma}$.  The problem with this notation is that it does not represent the spirit of the order independence axiom.  If it does not matter what order three or more parts are joined in, then it would be convenient if notation treated all the parts on an equal footing.   The tensorial notation accomplishes this though at the cost of assuming something about the nature of composition. This is the assumption of composition locality given below.

Let us write
\begin{equation}
\mathsf{ (A, B)_\alpha = A^{a_1}}[x] \mathsf{B_{a_1}}[y]
\end{equation}
where $\alpha=(\mathsf{a}, x,y)$ and $\mathsf a$ is the type of join, $x$ is the \lq\lq position" of the join to object $\mathsf A$, and $y$ is the \lq\lq position" of the join to object $\mathsf B$.  Here we are just creating the possibility that joins, such as $\alpha$, can be separated into type information and position information (since this is the case in the examples we will consider).  However, we don't have to do this. We could simply have a new type for every pair of positions.  Hence, this is just notation (we are not assuming anything extra quite yet).  The integer label, 1, on the right is not strictly necessary here but is essential when we have more than one join in the tensorial notation.  Now, if we have three objects joined together, we have
\begin{equation}\label{examplewithpositions}
\mathsf{ ((A, B)_\alpha, C)_\beta} = (\mathsf{A^{a_1}}[x] \mathsf{B_{a_2}}[y])^{\mathsf{d_2}}[u] \ \mathsf{C_{d_2}}[v]
\end{equation}
where $\alpha=(\mathsf{a}, x, y)$ and $\beta=(\mathsf{d}, u,v)$.   The bracketing means that we do not, strictly, need the integer labels. However, we include them as we are taking steps to the tensorial notation where we will need them.
Recall that the particular integers used are of no significance - they are just labels.  The example in (\ref{examplewithpositions}) can be read as first joining $\mathsf A$ at $x$ to $\mathsf B$ at $y$ by join $\mathsf a$ and then joining the composite object $\mathsf{ (A, B)_\alpha}$ at $u$ to $\mathsf C$ at $v$ to by join $\mathsf d$.  It seems reasonable that when we join $\mathsf C$ in this way we can understand the join $\mathsf d$ to be composed of a join to $\mathsf A$ and a join to $\mathsf B$.  Indeed, consider the object
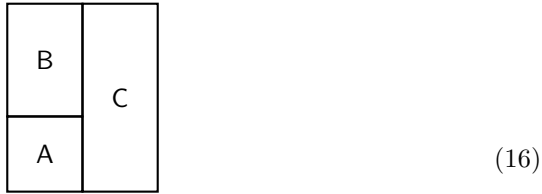
\begin{equation}
\begin{Compose}{0}{0}
\Csquare{A}{2}{0,0} \Crectangle{B}{2}{3}{0,5} \Crectangle{C}{2}{5}{4,3}
\end{Compose}
\end{equation}
Here a bigger rectangle is composed from a square and two smaller rectangles.  We can explode this to illustrate where the joins are
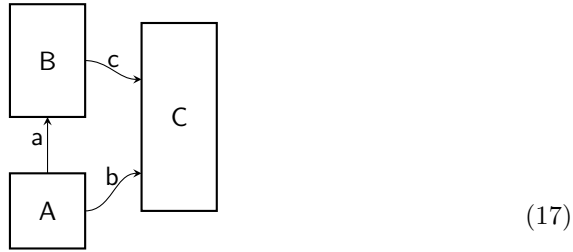
\begin{equation}\label{tripartiteexample}
\begin{Compose}{0}{0}
\Csquare{A}{2}{0,0} \Crectangle{B}{2}{3}{0,8} \Crectangle{C}{2}{5}{7,5}
\jointb{A}{0}{B}{0} \csymbol[-15,-5]{a}
\joinrl{A}{0}{C}{-3} \csymbol[-2,10]{b}
\joinrl{B}{0}{C}{2} \csymbol{c}
\end{Compose}
\end{equation}
In this example $\mathsf C$ is joined separately to $\mathsf A$ and $\mathsf B$. We can think of the join, $(\mathsf d, u, v)$ in (\ref{examplewithpositions}) as being equal to $(\mathsf{bc}, qr, st)$ where $\mathsf A$ at $q$ is joined to $\mathsf C$ at $s$ by a join of type $\mathsf b$ and $\mathsf B$ at $r$ is joined to $\mathsf C$ at $t$ by a join of type $\mathsf c$.
We would like our notation to reflect this structure.  We will first simplify our notation by absorbing the positions of the joins into the specification of the object.  Thus, rather than writing $\mathsf{ A^{a_1b_2}}[xq]$ we will write $\mathsf{A^{a_1b_2}}$ where the positions $x$ and $q$ have been absorbed into the definition of $\mathsf A$.  Similarly, we write $\mathsf{ B_{a_1}^{c_3}}$ and $\mathsf{C_{b_2c_3}}$ absorbing the relevant positions into the definitions of $\mathsf B$ and $\mathsf C$ respectively.  It may, sometimes, be useful to go back to the more cumbersome notation where we explicitly give the positions.

With this notation in place we can provide a new axiom that is motivated by the above discussion.
\begin{description}
\item[Composition locality]
We can represent a multipartite composite object as follows:
\begin{equation}\label{tripartitecomploc}
\mathsf{
A^{a_1b_2} B^{c_3}_{a_1} C_{b_2c_3} }
\end{equation}
for a tripartite object, and
\begin{equation}
\mathsf{
A^{a_1b_2c_3} B^{d_4e_5}_{a_1} C^{f_6}_{b_2d_4} D_{c_3e_5f_6}   }
\end{equation}
for an object with four components, and so on.  In such expressions the order of the objects and the particular choice of integer labels are unimportant. For example,
\begin{equation}\label{orderunimportant}
\mathsf{ A^{a_1c_3} B^{d_4}_{a_1} C_{c_3d_4} = A^{a_7c_3} C_{c_3d_1} B^{d_1}_{a_7} = B^{d_4}_{a_1} A^{a_1c_5} C_{c_5d_4} = \dots 
}
\end{equation}
In each of these equivalent expressions we maintain the same types of joins between objects.
\end{description}
The notation in (\ref{tripartitecomploc}) exactly captures the structure of the diagram in (\ref{tripartiteexample}).  We can make the diagram a little more abstract. Thus, we can represent the compositional structure shown in (\ref{tripartiteexample}) by
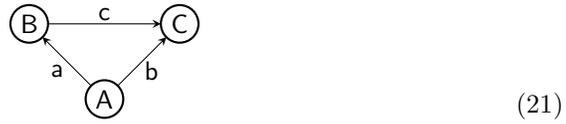
\begin{equation}
\begin{Compose}{0}{0}
\Ucircle{A}{0,0} \Ucircle{B}{-4,4} \Ucircle{C}{4,4}
\joincc[below left]{A}{135}{B}{-45} \csymbol{a}
\joincc[below right]{A}{45}{C}{-135} \csymbol{b}
\joincc{B}{0}{C}{180} \csymbol{c}
\end{Compose}
\end{equation}
Indeed, since one can use diagrammatic notation, this diagram and this symbolic notation are just two ways of notating the same thing.  Likewise, when we have four parties, we have symbolic and diagrammatic notation:
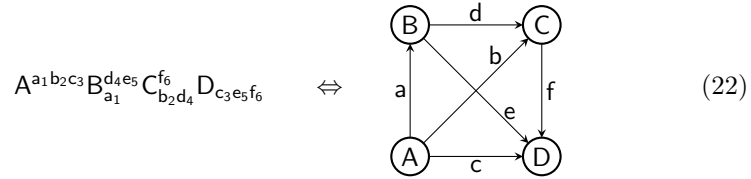
\begin{equation}
\mathsf{
A^{a_1b_2c_3} B^{d_4e_5}_{a_1} C^{f_6}_{b_2d_4} D_{c_3e_5f_6}   }   ~~~~~\Leftrightarrow~~~~
\begin{Compose}{0}{-1}
\Ucircle{A}{0,0} \Ucircle{B}{0,7} \Ucircle{C}{7,7} \Ucircle{D}{7,0}
\joincc[left]{A}{90}{B}{-90} \csymbol{a} \joincc[above right]{A}{45}{C}{-135} \csymbol[15,40]{b} \joincc[below]{A}{0}{D}{180} \csymbol{c}
\joincc[above]{B}{0}{C}{180} \csymbol[0,5]{d} \joincc[below right]{B}{-45}{D}{135} \csymbol[37,-20]{e}
\joincc[right]{C}{-90}{D}{90} \csymbol{f}
\end{Compose}
\end{equation}
The composition locality axiom enables us to represent composite objects in this way.  It is worth commenting on what is ruled out by this axiom.  It could be the case that when we join two objects, $\mathsf A$ and $\mathsf B$, some new joins become possible that cannot be understood in terms of joins to $\mathsf A$ and $\mathsf B$ separately.  It could be the case that some kinds of join can only be regarded as being between more than two objects.  It could matter what order we understand the composition of the objects to be taken in. These things are ruled out if we assume composition locality.  In the symbolic notation this is illustrated by the fact that we can write the objects in any order (as illustrated in (\ref{orderunimportant})).  In the diagrammatic notation it is illustrated by the fact that we can place the objects anywhere on the page (so long as we maintain the joins between the objects).

In general joins are asymmetric. This is why we use asymmetric notation.  For example, we place glue on the object with raised superscript before pushing it into place on the object with lowered superscript.  Any asymmetrically described join can be described in the other direction. Thus, we may apply glue on the object with lowered superscript object then press it into place on the object with raised superscript.  We denote this reverse description of a join by using a ${\mathnormal R}$ superscript. So, if $\mathsf a$ is some join type, then $\mathsf{a}^{\mathnormal R}$ is the same join described in the reverse direction.  This means that
\begin{equation}
\mathsf{
A^{a_1} B_{a_1}= A_{a^{\mathnormal R}_1} B^{a^{\mathnormal R}_1}  }
\end{equation}
Hence, definitionally, we can raise and lower superscripts and subscripts in this case if we append the $R$ superscript.  If we have more subscripts and superscripts we would like to be able to do the same.  Hence we assume
\begin{description}
\item[$R$-enablement.]  In any description of a composite object in tensorial notation we can reverse the direction of any particular join by changing a subscript to a superscript, changing the corresponding superscript to a subscript and appending a $R$ to each. For example,
\[ \mathsf{ A^{a_1b_2} C_{a_1}^{a_3} B_{b_2a_3} = A^{{\ } b_2}_{a_1^{\mathnormal R}} C^{a^{\mathnormal R}_1a_3} B_{b_2a_3}  }  \]
\end{description}
We note that, when subscripts are raised or lowered we should maintain a record of the positions of the joins.  We can do this by appropriate indentation of the labels (though, for the most part, we will not worry about this in this paper).
The assumption of $R$-enablement is very natural.  In diagrammatic notation this simply corresponds to reversing the direction of the arrows.  In diagrammatic notation we reverse the direction of the join simply by reversing the direction of the arrows.  For example,
\begin{equation}
\begin{Compose}{0}{-0.5}
\Ucircle{A}{0,0} \Ucircle{B}{-4,4} \Ucircle{C}{4,4}
\joincc[below left]{A}{135}{B}{-45} \csymbol{a}
\joincc[below right]{A}{45}{C}{-135} \csymbol{b}
\joincc{B}{0}{C}{180} \csymbol{c}
\end{Compose}
~~=~~
\begin{Compose}{0}{-0.5}
\Ucircle{A}{0,0} \Ucircle{B}{-4,4} \Ucircle{C}{4,4}
\joincc[below left]{B}{-45}{A}{135} \csymbol[-6,-6]{a^{\mathnormal R}}
\joincc[below right]{A}{45}{C}{-135} \csymbol{b}
\joincc{C}{180}{B}{0} \csymbol[0,8]{c^{\mathnormal R}}
\end{Compose}
\end{equation}
It is natural to lump together the last two axioms.  If we have $R$-enablement and composition locality then we will say that we have $R$-enabled composition locality. In this case we can use the tensorial notation and reverse the direction of any join.

We will now show that $R$-enabled composition locality implies the order independence axiom.   We can write
\begin{equation}
\mathsf{
A^{a_1b_2} B^{c_3}_{a_1} C_{b_2c_3} = (A, B^{c_3} C_{c_3})_\alpha  }
\end{equation}
where $\alpha=(\mathsf{ab}, xy, uz)$ corresponds to joining $\mathsf A$ at $xy$ to $\mathsf{ B^{c_3} C_{c_3} }$ at $uz$ by a join of type $\mathsf{ab}$ (where $u$ is a position at $\mathsf B$ and $z$ is a position at $\mathsf C$).   It is then clear that
\begin{equation}
\text{$R$-enabled composition locality} \Rightarrow \text{the order independence axiom}
\end{equation}
To see this we write
\begin{equation}
\mathsf{
A^{a_1b_2} B^{c_3}_{a_1} C_{b_2c_3} = A^{b_2}_{a^{\mathnormal R}_1} B^{a^{\mathnormal R}_1c_3} C_{b_2c_3} = (B, A^{b_2} C_{b_2})_\beta
}
\end{equation}
where $\beta=(\mathsf{a^{\mathnormal R} c}, uv, xw)$ corresponds to joining $\mathsf B$ at $uv$ to $\mathsf A$ at $x$ and $\mathsf C$ at $w$ by a join of type $\mathsf a^{\mathnormal R} c$.

We have not proven that the order independence axiom implies the composition locality axiom.  It would be interesting to find examples of composite objects that satisfy the very fundamental order independence axiom while violating composition locality.  In this case we could not use the tensorial notation.

\subsection{Composition Locality and Null Joins}\label{SecMoreonNulljoins}

Some properties follow from the uniqueness axiom for null joins.  Thus, consider two joins, $\alpha=(\mathsf a, x, y)$ and $\beta=(\mathsf b, u, v)$ where both of these joins are, actually, the null join.  By uniqueness, $\alpha=\beta= 0$. Hence, we have $\mathsf{a=b}$.   What this means is that the join type is the same for any null join between a given pair of objects. We will denote the null join type by the symbol $\mathsf 0$.  Thus, in this case we have $\mathsf a=\mathsf b = \mathsf 0$.  We also have $x=u$ and $y=v$.  This means that the null join does not really have a position at either object.   It is also worth noting that $\mathsf{ 0^{\mathnormal R}= 0}$ by the symmetry axiom for null joins.  All these properties are consistent with the idea that the null join just corresponds to taking two objects to be part of the picture without specifying any relationship beyond this.

Consider the composite object $\mathsf {(A, B^{a_1}C_{a_1})_0 }$.  This consists of two disjoint parts (that is parts connected by the null join).  According to the $R$-enabled composition locality axiom we can write
\begin{equation}
\mathsf {  (A, B^{a_1}C_{a_1})_0   = A^{b_2c_3} B^{a_1}_{b_2} C_{a_1c_3} = B^{a_1b_2^{\mathnormal R}} A^{c_3}_{b_2^{\mathnormal R}} C_{a_1 c_3} = ( B, A^{c_3} C_{c_3})_\alpha
= ((A,C)_\beta, B)_{\alpha^{\mathnormal R}}
}
\end{equation}
for some $\mathsf b$, $\mathsf c$, $\beta$ and $\alpha$.  Now, according to the refinement axiom for null joins, we must have $\beta=0$ and hence $\mathsf{ c=0}$.  By a similar argument, we must have $\mathsf{ b=0}$.  Hence,
\begin{equation}
\mathsf {  (A, B^{a_1}C_{a_1})_0   = A^{0_20_3} B^{a_1}_{0_2} C_{a_10_3} }
\end{equation}
We can notate this this simply by omitting the null joins. Thus, we can write
\begin{equation}
\mathsf {  (A, B^{a_1}C_{a_1})_0   = A B^{a_1} C_{a_1}   }
\end{equation}
where the null joins are taken to be implicit.  Another example is the following
\begin{equation}\label{fourwithnulljoins}
\mathsf{
(A^{a_1} B_{a_1},C^{b_2} D_{b_2} )_0 = A^{a_1} B_{a_1} C^{b_2} D_{b_2}
}
\end{equation}
Again, the null joins are taken to be implicit.   We can show that the right of (\ref{fourwithnulljoins}) follows from composition locality and the refinement axiom for null joins.   This is true in general. Every pair of objects that are not joined by some non-null join are joined by a null join.   Hence, when we adopt these axioms, we are effectively forced to notate null joins this way in the tensorial notation (there is no point in explicitly including the null joins as it is clear where must be).  We can also see, immediately, that if we do notate null joins in this way then all the null join axioms are satisfied.  Diagrammatically we have
\begin{equation}
\mathsf{  A^{a_1} B_{a_1} C^{b_2} D_{b_2} } ~~~ \Leftrightarrow ~~~
\begin{Compose}{0}{-1}
\Ucircle{B}{0,6} \Ucircle{C}{5, 6}
\Ucircle{A}{0,0} \Ucircle{D}{5,0}
\joincc[left]{A}{90}{B}{-90} \csymbol{a}
\joincc[right]{C}{-90}{D}{90} \csymbol{b}
\end{Compose}
\end{equation}
Thus, disjoint parts of a composite object are naturally represented by disjoint parts of the graph.

\subsection{Pruning}

We now wish to address the possibility that we may have more joins listed in specifying an object than necessary.  We may be able to prune the graph while still describing the composite object as fully as we require for our purposes.  It is worth setting up this discussion with an example.

The notion of a join is really an abstraction.  It does not have to correspond to two objects actually being in contact.  For example, we may have a join type, $\mathsf b$, between two squares (with unit length edges) that corresponds to \lq\lq joining" the second square one unit of distance to the right of the first square.  Thus our object looks like
\begin{equation}\label{twosquaresoneunit}
\begin{Compose}{0}{0}
\Csquare{A}{2}{0,0} \Csquare{C}{2}{8,0}
\linebyhand[thick, ->]{2, 0}{6,0} \csymbolalt[0,20]{1\text{unit}}
\end{Compose}
\end{equation}
which we can write as
\begin{equation}
\mathsf{A^{b_1} C_{b_1} }~~~\Leftrightarrow~~~
\begin{Compose}{0}{0}
\Ucircle{A}{0,0}  \Ucircle{C}{6,0}
\joincc[above]{A}{70}{C}{110} \csymbol[0,7]{b}
\end{Compose}
\end{equation}
This join does not correspond to placing the two squares exactly next to each other but letting them have a given displacement from one another.   Now, if we place a third square directly between the two objects then we have the following object:
\begin{equation}
\begin{Compose}{0}{0}
\Csquare{A}{2}{0,0}\Csquare{B}{2}{4,0} \Csquare{C}{2}{8,0}
\end{Compose}
\end{equation}
If the join type corresponding to placing the second square immediately to the right of the first square is denoted $\mathsf a$ then we can denote this composite object as
\begin{equation}
\mathsf{ A^{a_1b_3} B_{a_1}^{a_2} C_{a_2b_3} } ~~~ \Leftrightarrow~~~
\begin{Compose}{0}{0}
\Ucircle{A}{0,0} \Ucircle{B}{5,0} \Ucircle{C}{10,0}
\joincc[above]{A}{0}{B}{180} \csymbol{a} \joincc[above]{B}{0}{C}{180} \csymbol{a} \joincc[above]{A}{70}{C}{110} \csymbol[0,7]{b}
\end{Compose}
\end{equation}
If we had also to include joins where the squares had relative displacements of two units, three units, and so on, then as the number of squares increase the number of joins would increase faster ($N(N-1)$ joins are possible for $N$ objects).  Fortunately, we note that the fact that $\mathsf A$ and $\mathsf C$ are joined by a join of type $\mathsf c$ is implied by the other joins (the fact that $\mathsf A$ is joined to $\mathsf B$ and then $\mathsf B$ is joined to $\mathsf C$ by joins of type $\mathsf a$).   Hence, we do not lose any information by pruning the description to the form
\begin{equation}
\mathsf{ A^{a_1} B_{a_1}^{a_2} C_{a_2} } ~~~ \Leftrightarrow~~~
\begin{Compose}{0}{0}
\Ucircle{A}{0,0} \Ucircle{B}{5,0} \Ucircle{C}{10,0}
\joincc[above]{A}{0}{B}{180} \csymbol{a} \joincc[above]{B}{0}{C}{180} \csymbol{a}
\end{Compose}
\end{equation}
However, this is a little unsatisfactory.  If a third object ($\mathsf B$) is present then we can prune join $\mathsf c$. Otherwise we cannot.  In this example, we can avoid this problem by enlargening our set of objects to include \lq\lq empty space" objects.  Thus, we denote by $\mathsf E$ an empty unit square.  For definiteness, we can imagine that the squares are made out of sheet aluminium.  Then an empty square means that we have a one unit area with no sheet aluminium in it.  Then we can denote the object illustrated in (\ref{twosquaresoneunit}) by the \lq\lq pruned" notation
\begin{equation}
\mathsf{ A^{d_1} E_{d_1}^{e_2} C_{e_2} } ~~~ \Leftrightarrow~~~
\begin{Compose}{0}{0}
\Ucircle{A}{0,0} \Ucircle{E}{5,0} \Ucircle{C}{10,0}
\joincc[above]{A}{0}{B}{180} \csymbol[0,4]{d} \joincc[above]{E}{0}{C}{180} \csymbol{e}
\end{Compose}
\end{equation}
where $\mathsf d$ and $\mathsf e$ denote the appropriate join types here.  If we proceed in this way then the number of joins need not grow as fast with the number of objects.  Further more, we can specify the object using only joins that correspond to the components being in contact.

This particular example motivates the following definition for the general case.
\begin{quote}
\emph{A sufficient set of join types} for some given set of objects is a set of join types such that any composite object formed from the given set of objects is fully specified (in the tensorial notation) if all joins of the type in the sufficient set are given where they exist between the components.
\end{quote}
This, in turn, motivates the following definition
\begin{quote}
\emph{A minimal set of join types} for some given set of objects is a sufficient set of join types having the property that, if any element is removed, we no longer have a sufficient set.
\end{quote}
The key idea behind sufficient (and minimal) sets of join types is that some joins, while they may exist, need not be specified in the context of a given set of objects. If we have a sufficient or, better, a minimal set of join types then we can proceed more efficiently.

\subsection{Boundaries}

Typically objects are joined to each other at boundaries of those objects. These boundaries live in space, time, or space-time.  An object will have a boundary that delimits what else it can be joined to.  This motivates the following axiom:
\begin{description}
\item[Boundary axiom:] given some sufficient set of joins, $\mathcal{J}_\text{suff},$ then for any object $\mathsf A$ and any join $\mathsf a\in\mathcal{J}_\text{suff}$ admitted by $\mathsf A$ there exists a unique join $\mathsf b$ in $\mathcal{J}_\text{suff}$ such that if $\mathsf A$ is joined to some object by $\mathsf{ab}$ then no further joins in $\mathcal{J}_\text{suff}$ at $\mathsf A$ are possible other than the null join.
\end{description}
Let us call the unique join, $\mathsf b$, in this definition the \emph{complement to} $\mathsf a$ \emph{for} $\mathsf A$.  We can write it as $\mathsf{\bar a_A}$.  By $\bar 0_\mathsf{A}$ we denote the complement to the null join.  This deserves a special name so we call it the complete join for $\mathsf A$.  If $\mathsf A$ is joined to any other object by its complete join then it does not admit further joins (other than the null join).  Thus, the complete join can be thought of as representing the boundary of $\mathsf A$.

Once we have the boundary axiom, we can introduce a final refinement to the tensorial notation for composition.  We can denote an object by $\mathsf{A^{a_1}_{b_2}}$ where we demand that $\mathsf{ab^{\mathnormal R}=\bar 0_A}$.  The subscripts and superscripts may be composite. For example, we may have $\mathsf{A^{a_1b_2}_{c_3d_4}}$ where $\mathsf{abc^{\mathnormal R} d^{\mathnormal R}=\bar 0_A}$.  The advantage of this notation is that if there are joins left open we can read it off. For example, we may have a composite object
\begin{equation}
\mathsf{A^{a_1b_2}B^{c_3}_{a_1} C_{c_3d_4} }
\end{equation}
This has open joins $\mathsf{b}$ at $\mathsf A$ and $\mathsf d^{\mathnormal R}$ at $\mathsf C$.  This is even clearer if we represent this diagrammatically
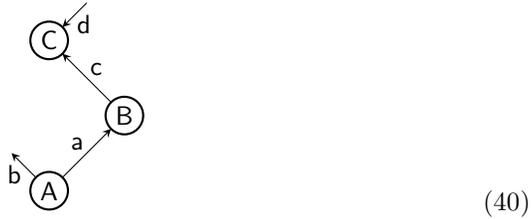
\begin{equation}
\begin{compose}
\Ucircle{A}{0,0}\Ucircle{B}{4,4} \Ucircle{C}{0,8}
\joincc[above left]{A}{45}{B}{-135} \csymbol{a}
\joincc[above right]{B}{135}{C}{-45} \csymbol{c}
\thispoint{b2}{-2,2} \joincc[below left]{A}{135}{b2}{-45} \csymbol{b}
\thispoint{d4}{2,10} \joincc[below right]{d4}{-135}{C}{45} \csymbol{d}
\end{compose}
\end{equation}

\section{The Composition Principle}\label{SecCompositionprincple}

Up to now we have discussed the \emph{description} of composite objects and we have developed the tensorial notation for this description.  In physics we are also interested in predicting the values of properties pertaining to the given physical object.  Here are a few examples of properties we might try to predict.
\begin{description}
\item[Possible or impossible.]  We may be interested in whether some particular object is actually possible.  If it is possible we can return the value 1 and if not, we can return the value 0.  An example of this are what we might call \lq\lq Penrose objects" consisting of beams with square cross section that are joined at right angles at their ends.  The Penrose triangle is an example of an impossible object and so we should return the value 0.  The possible/impossible approach is a way of understanding deterministic theories (such as Newtonian dynamics).  Processes that violate the predictions of the theory are, within the context of this theory, impossible.
\item[Probability.]  We may ask what the probability of a particular set of outcomes is.  An example where we do this is for probabilistic circuits.
\item[Dimensions.]  We may be interested in what the dimensions of some object is (its height, width, length).
\item[Minimum distances.] We may be interested in the minimum distance between points in the object (as measured within the object).
\end{description}
In a typical situation in physics we are only interested in some subset of properties (for example, in thermodynamics, we are interested in the values of certain macroscopic variables but we are not interested in the velocity of individual atoms).   We define the following notion
\begin{quote}
{\bf The generalized state} is a mathematical object, $A$,  associated with an object, $\mathsf A$, which can be used to calculate the value of those properties we are interested for this object.
\end{quote}
Typically in physics a state pertains to a given time and is used to make predictions for later times.  The generalized state is a more general notion than this since we may be making predictions of a more general type (such as in the examples given above).

A key question is how do we calculate the generalized state for a composite object? We propose the following principle.
\begin{quote}
{\bf THE COMPOSITION PRINCIPLE:} The generalized state for a composite object can be calculated from the generalized states for the components by means of a calculation having the same structure as the description of the composition of that object.
\end{quote}
For example, we can write
\[  \mathsf{ A^{a_1b_2} B_{a_1}^{c_3} C_{b_2c_3}} ~~\text{has generalized state}~~~  A^{a_1b_2} B_{a_1}^{c_3} C_{b_2c_3} \]
where $A^{a_1b_2}$, $B_{a_1}^{c_3}$, and $C_{b_2c_3}$ are the generalized states associated with $\mathsf{A^{a_1b_2}}$, $\mathsf{B_{a_1}^{c_3}}$, and $\mathsf{C_{b_2c_3}}$ respectively.  The subscript/superscript placement corresponds to some mathematical operations respecting the composite structure.  We will not show that the composition principle holds for all physical situations (such as those given).  There may be counterexamples. However, we will illustrate the principle with a few simple examples.

\subsection{Circuits}

Here is an example of the composition principle in action.  We consider circuits as described in Sec.\ \ref{SecCircuits} where we are interested in calculating the probability of seeing outcomes at the operations that are in the associated outcome sets.  A much discussed assumption for probabilistic circuit models is \emph{tomographic locality} (see \cite{hardy2011reformulating} and references therein).  This property has many equivalent formulations.  The most common is that the state associated with a bipartite system can be determined by local measurements on each of the systems.  In the case where we have \emph{tomographic locality} the composition principle holds for calculating probabilities.   To calculate the probability of the circuit
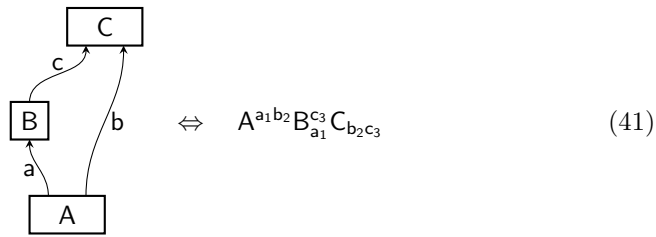
\begin{equation}
\begin{Compose}{0}{-1.3}
\Crectangle{A}{2}{1}{0,0} \Crectangle{B}{1}{1}{-2,5} \Crectangle{C}{2}{1}{2,10}
\jointb[left]{A}{-1}{B}{0} \csymbol[0,-4]{a} \jointb{B}{0}{C}{-1} \csymbol{c} \jointb[right]{A}{1}{C}{1} \csymbol[5,0]{b}
\end{Compose}
~~~ \Leftrightarrow~~~ \mathsf{ A^{a_1b_2} B_{a_1}^{c_3} C_{b_2c_3}}
\end{equation}
we can write
\[ \text{Prob}( \mathsf{ A^{a_1b_2} B_{a_1}^{c_3} C_{b_2c_3}}) =  A^{a_1b_2} B_{a_1}^{c_3} C_{b_2c_3}  \]
where $A^{a_1b_2}$, $B_{a_1}^{c_3}$, and $C_{b_2c_3}$ are tensors. The subscript/superscript placement now corresponds to Einstein summation.   In this example $A^{a_1b_2} B_{a_4}^{c_3}$ corresponds to \emph{multiplying} two numbers (the elements of the tensors with these particular values of the subscripts and superscripts) and then the repeated index corresponds to summation.  Thus, we have two operations, multiplication and summation. This means that something like $C^{a_1} D_{a_1}$ is a sum of products.  These two operations guarantee that we can do the calculation in any order.  Thus, we can write
\[ A^{a_1b_2} B_{a_1}^{c_3} C_{b_2c_3} = C_{b_2c_3} A^{a_1b_2} B_{a_1}^{c_3} = B_{a_1}^{c_3} A^{a_1b_2} C_{b_2c_3} = \dots \]
by the associativity property of multiplication and we can perform the three different summations in any order by the associativity property of summation.  The composition principle is satisfied since, for the purpose of calculating probabilities, we have generalized states (the tensors) and the are determined by means of a calculation having the same structure as compositional structure.

Quantum theory satisfies the principle of tomographic locality and it can be formulated in such a way that the generalized state is given by tensors as just described. However, for quantum theory, there exists a more tailor-made way of representing the generalized state in such a way that the composition principle is satisfied.  This is to represent the generalized state of the operation $\mathsf{B_{a_1}^{c_3}}$, for example, by an hermitian operator $\hat B_\mathsf{a_1}^\mathsf{c_3}$ satisfying certain properties (see \cite{hardy2011reformulating,hardy2012operator} for more details).  Then we can write
\[ \text{Prob}( \mathsf{ A^{a_1b_2} B_{a_1}^{c_3} C_{b_2c_3}}) =  \hat A^\mathsf{a_1b_2} \hat B_\mathsf{a_1}^\mathsf{c_3} \hat C_\mathsf{b_2c_3}  \]
In the notation on the right, the repeated label indicates partial trace over the appropriate part of the operator space (as outlined in \cite{hardy2011reformulating,hardy2012operator}).

Tomographic locality has been used as a postulate in many of the recent reconstructions of quantum theory.  It is intriguing that it is connected to the even deeper idea of the composition principle.  In the context of probabilistic circuit theories, the composition principle follows from tomographic locality.  It is possible that the converse is true.  This raises the intriguing possibility of using the composition principle as a basis for reconstructing quantum theory.  The composition principle is a deeper principle than tomographic locality and it may play a role in formulating physics that goes beyond the circuit setting of quantum theory (such as a theory of quantum gravity).

\subsection{Labeled Tiles}

We consider square tiles of unit length that are labeled $n=1, 2, \dots$.  Let the $n$th such tile be $\mathsf T[n]$.  A complete set of join types is $\{\mathsf{ x, y, 0} \}$.  The join type $\mathsf x$ corresponds to placing one immediately to the right of the other as follows
\begin{equation}
\mathsf{ T^{x_1}}[m]\mathsf{ T_{x_2}}[n] ~~~~ \Leftrightarrow~~~~
\begin{Compose}{0}{0}
\usquare{m}{0,0} \csymbolalt{m} \usquare{n}{2,0} \csymbolalt{n}
\end{Compose}
\end{equation}
The join type $\mathsf y$ corresponds to placing one tile immediately above the other
\begin{equation}
\mathsf{ T^{y_1}}[m]\mathsf{ T_{y_2}}[n] ~~~~ \Leftrightarrow~~~~
\begin{Compose}{0}{0}
\usquare{m}{0,0} \csymbolalt{m} \usquare{n}{0,2} \csymbolalt{n}
\end{Compose}
\end{equation}
The third join type is the null join where we simply consider the two tiles as part of the same picture without specifying their relationship any further.
\begin{equation}
\mathsf{ T}[m]\mathsf{ T}[n] ~~~~ \Leftrightarrow~~~~
\begin{Compose}{0}{0}
\usquare{m}{0,0} \csymbolalt{m} \usquare{n}{5.5,2.7} \csymbolalt{n}
\end{Compose}
\end{equation}
We can define $\mathsf x^{\mathnormal R}$ and $\mathsf y^{\mathnormal R}$ to correspond to joins in the opposite direction.  For example,
\begin{equation}
\mathsf{ T^{x^{\mathnormal R}_1}}[m]\mathsf{ T_{x^{\mathnormal R}_2}}[n] ~~~~ \Leftrightarrow~~~~
\begin{Compose}{0}{0}
\usquare{n}{0,0} \csymbolalt{n} \usquare{m}{2,0} \csymbolalt{m}
\end{Compose}
\end{equation}
We can build up highly composite objects by joining many tiles.  For example,
\begin{equation}\label{objectD}
\mathsf{D=T^{x_2}_{y_1}[1] \ T^{y_1x_3}[2] \ T_{x_3}^{y_4}[3] \ T_{y_3x_2}^{y_5}[4]  \ T_{y_5}[5] \ T^{x_6}[6] \ T_{x_6}[7]  } ~~ \Leftrightarrow ~~
\begin{Compose}{0}{0}
\Usquare{1}{0,0} \Usquare{2}{0,-2} \Usquare{3}{2,-2} \Usquare{4}{2,0} \Usquare{5}{2,2} 
%
\Usquare{6}{6.4,0.7}\Usquare{7}{8.4,0.7}
\end{Compose}
\end{equation}
This particular composite object is disjoint (it consists of two disjoint parts).

We are interested in calculating geometric properties having to do with the relative displacement of tiles (where defined).  Hence, we define the following generalized state for objects that have no disjoint parts
\begin{equation}
A=\{ \big((m,n), (\Delta x,\Delta y)\big): \text{for all tile labels}\ m,n\ \text{in}\ \mathsf A\}
\end{equation}
where $\Delta x$ is the horizontal displacement from tile $m$ to tile $n$ and $\Delta y$ is the vertical displacement between these tiles (so $(\Delta x,\Delta y)$ is the displacement between these two tiles).  We can obtain the generalized state for objects that do have disjoint parts by using the null join.  Corresponding to the object, $\mathsf {A B}$ we have
\begin{equation}
A B := A \cup B
\end{equation}
(recall that we suppress the $\mathsf 0$ but we could write $\mathsf{A^{0_1} B_{0_1} }$ for this object).
For the composite object $\mathsf D$ in (\ref{objectD}) we would get all the displacements in each of the two disjoint parts.  However, no displacements would be specified between the two disjoint parts since these are not defined.

We can join two composite objects by an $\mathsf x$ join by specifying the tiles in each object where this join is to occur. So,
\begin{equation}
\mathsf{C}=\mathsf{A^{x_1}}[u] \mathsf{B_{x_1}}[v]
\end{equation}
means that we join $\mathsf A$ at tile $u$ to $\mathsf B$ at tile $v$ by an $\mathsf x$ type join.  The associated generalized state is given by
\begin{equation}
C= A^{x_1}[u] B_{x_1}[v] = A \cup B \cup \mathcal{X}(A[u],B[v])
\end{equation}
where
\begin{multline}
\mathcal{X}(A[u],B[v])= \\
\{ \big( (m, n), {\bf \Delta}_{(m,u)}+(1,0)+{\bf \Delta}_{(v,n)} \big): \forall  (m, u)\in \mathsf{A}\ \text{and}\ (v,n) \in \mathsf{B} \}
\end{multline}
and ${\bf \Delta}_{(m,u)}$ is the displacement between tile $m$ and tile $u$.  This is the set of displacements between $\mathsf A$ and $\mathsf B$ that are established by this new join.  For this join to be consistent no two tiles should end up at the same position.  Clearly, one can determine this by taking an appropriate function of $C$.  The effect of a $\mathsf y$ join is given by
\begin{equation}
E= A^{y_1}[u] B_{y_1}[v] = A \cup B \cup \mathcal{Y}(A[u],B[v])
\end{equation}
where
\begin{multline}
\mathcal{Y}(A[u],B[v])= \\
\{ \big( (m, n), {\bf \Delta}_{(m,u)}+(0,1)+{\bf \Delta}_{(v,n)} \big): \forall  (m,u)\in \mathsf{A}\ \text{and}\ (v,n) \in \mathsf{B} \}
\end{multline}
If we just have a single tile, $\mathsf T[n]$, then the corresponding generalized state is
\begin{equation}
T[n]= \{ ((n,n), (0,0)) \}
\end{equation}
It is now easy to see that, in building up the generalized state corresponding to any composite object (such as $\mathsf D$ in (\ref{objectD})), we get the same answer no matter what order we do the calculation.  Further, it is clear that geometric properties are all given by appropriate functions of the resulting generalized state.  The composition principle clearly holds since we obtain generalized states by means of a calculation having the same structure as that of the description of the composite object.

When we have more than one join we have an additional consistency condition.  Namely, the displacement between any pair of tiles should be the same.  If the joins are inconsistent then we will get more than one displacement for any given pair of tiles.  This consistency condition is easily checked by taking an appropriate function of the mathematical object corresponding to the constructed composite object.

It is worth adding that we could have a more compact specification of the generalized state. Thus, rather than specifying the relative displacement of every pair of tiles in each disjoint part, we could simply specify the relative displacement of every tile from a given \emph{fiducial} tile in each disjoint part.  We could expand this compressed information simply by subtraction to get the displacement between any two tiles.  A similar compression appears in the circuit framework above where we effectively specify a state by listing the probabilities for a fiducial set of measurement outcomes from which all others can be calculated.

\subsection{Other Examples}

It is a simple matter to set up the theory for other examples but lack of space prohibits us from doing so here.  For example, Penrose objects can be specified by calculating the displacement between the two ends in the coordinate system that is given by the average of coordinate systems for each beam.  Arbitrary regions of spacetime with fields can be joined by requiring that the fields match and a certain number of derivatives match at the boundaries.  Given a new example it requires a certain amount of ingenuity to set up the appropriate mathematical objects along with appropriate joining conditions.  However, it seems likely that this can be done for any physical example.  This raises the question of whether assuming the composition principle actually amounts to assuming something of the world, or whether it can always be made to be true.  Even if the latter is the case, the composition principle could be a useful tool in constructing new physical theories.  In particular, in a theory of quantum gravity we must, presumably, build up a picture of bigger things by joining smaller things. However, we can expect to have indefinite causal structure and consequently boundaries between such smaller things can be expected to be causally fuzzy.  In this case, the new tools developed in this paper could be very useful for the process of theory construction.

\section{Conclusions}

We have developed the tensorial notation for describing composite objects and explored the assumptions going into this notation using the more primitive bipartite notation.  In describing objects we are not making any predictions for them.  The composition principle suggests that there is a correspondence between objects and certain mathematical objects (generalized states) that allow us to make predictions of properties of the objects by means of equations that have the same structure as the description of the composition of those objects.  We have given a few simple examples where the composition principle can be seen to hold.  We make the stronger claim that any reasonable physical theory can be formulated in a way that this principle holds. Further, we claim that all physical theories \emph{ought} to be formulated in a way that makes it clear that the composition principle holds.    This is a potentially very useful principle in constructing new physical theories, such as a theory of quantum gravity.

\section*{Acknowledgements}

Research at Perimeter Institute is supported by the Government
of Canada through Industry Canada and by the Province of Ontario through the Ministry
of Economic Development and Innovation.  This project was made possible in part through the support of a grant from the John Templeton Foundation. The opinions expressed in this publication are those of the author and do not necessarily reflect the views of the John Templeton Foundation.

\bibliography{quantumJan2013}
\bibliographystyle{plain}

\end{document}